\documentclass[preprint]{JHEP3} 

\usepackage{amsmath}

\newcommand{\cA}{{\cal A}}
\newcommand{\cD}{{\cal D}}
\newcommand{\cF}{{\cal F}}

\newcommand{\cN}{{\cal N}}
\newcommand{\cO}{{\cal O}}
\newcommand{\cQ}{{\cal Q}}
\newcommand{\cS}{{\cal S}}
\newcommand{\cU}{{\cal U}}
\newcommand{\cZ}{{\cal Z}}

\newcommand{\cAb}{{\overline{\cal A}}}
\newcommand{\cFb}{{\overline{\cal F}}}
\newcommand{\cDb}{{\overline{\cal D}}}
\newcommand{\cUb}{{\overline{\cal U}}} 

\newcommand{\Tr}{{\rm Tr\;}}
\newcommand{\hf}{\frac{1}{2}}
\newcommand{\qtr}{\frac{1}{4}}

\newcommand{\bn}{{\bf n}}
\newcommand{\hatbe}{\widehat{\boldsymbol e}}
\newcommand{\hatbmu}{\widehat{\boldsymbol \mu}}

\def\nn{\nonumber}
\def\bec{\begin{center}}
\def\eec{\end{center}}
\def\beq{\begin{equation}}
\def\eeq{\end{equation}}
\def\bea{\begin{eqnarray}}
\def\eea{\end{eqnarray}}

\title{Twisted supersymmetries in lattice $\cN=4$ super Yang-Mills theory}
\preprint{LA-UR-13-24390}

\author{Simon Catterall \\
Department of Physics, Syracuse University, Syracuse, NY 13244, USA}

\author{Joel Giedt \\
Department of Physics, Applied Physics and Astronomy, Rensselaer Polytechnic Institute, 110 8th Street, Troy NY 12065, USA}

\author{Anosh Joseph \\
Theoretical Division, Los Alamos National Laboratory, Los Alamos, NM 87545, USA}

\abstract{Recently it has been shown how a topologically twisted version of ${\cal N}=4$ 
super Yang-Mills may be discretized in such a way as to preserve one scalar supersymmetry 
at nonzero lattice spacing. The remaining fifteen supersymmetries are broken by terms of 
${\cal O}(a)$ where $a$ is the lattice spacing. One would like to know whether these remaining
 supersymmetries are regained in the continuum limit $a \to 0$ and, if not, how much tuning of 
the couplings in the lattice action is required. 
 In this paper we derive the form of these additional twisted supersymmetries
 by combining a set of discrete R-symmetries of the continuum theory 
with the action of the scalar supersymmetry.  We then argue that restoration 
of rotational symmetry in the continuum limit of the lattice theory 
likely implies restoration of R-symmetry and
hence should lead to an automatic enhancement to 
the full ${\cal N}=4$ supersymmetry without further fine-tuning.
}


\begin{document}

%########################################################
\section{Introduction}
\label{sec:intro}
%########################################################

In the recent past there has been a lot of progress in the area of lattice supersymmetry --- see \cite{Kaplan:2003uh, Giedt:2006pd, Catterall:2009it, Joseph:2011xy} for a set of reviews and references therein. The key to these new approaches lies in a reformulation of the continuum theory in terms of {\it twisted} variables. The process of twisting\footnote{The twisting technique leads to lattice theories which are essentially equivalent to those derived by orbifolding \cite{Unsal:2006qp, Catterall:2007kn}.} produces a nilpotent scalar supersymmetry $\cQ$ and an action that can be written, quite generally, as a sum of $\cQ$-exact and $\cQ$-closed pieces. It is this structure that allows one to write down a $\cQ$-invariant lattice theory \cite{Kaplan:2002wv, Cohen:2003xe, Cohen:2003qw, Kaplan:2005ta, Catterall:2003wd, Catterall:2001fr}. Only theories with sufficiently large numbers of supersymmetries can be twisted or orbifolded in this manner and thus implemented on a lattice. Among them, we are particularly interested in studying the well known theory, four-dimensional $\cN = 4$ supersymmetric Yang-Mills (SYM) and this is the focus of the current paper\footnote{Another approach to $\cN=4$ SYM using overlap fermions can be found in \cite{Elliott:2008jp}}. It is has been the subject of a few numerical studies \cite{Catterall:2012yq, Catterall:2011aa, Mehta:2011ud, Galvez:2012sv}, with more in progress.

It should be noted that other complementary Monte Carlo approaches have been proposed for studying
$\cN = 4$ SYM in the planar large $N$ limit see \cite{Ishii:2008ib, Ishiki:2008te,  Nishimura:2009xm, Ishiki:2009sg, Honda:2010nx, Hanada:2010kt, Hanada:2010gs}.  The lattice approach we discuss here, based on exact lattice supersymmetry, works for any gauge group, but retains only one supercharge. These other approaches work only in the large-$N$ limit, but retain larger numbers of exact supercharges. The results from the two approaches eventually should be compared in the region where they should coincide, namely at large $N$ and close to the continuum limit.

In addition, lattice formulations of two-dimensional SYM theories with four and eight supercharges which preserve  nilpotent supercharges are presented in \cite{Sugino:2003yb, Sugino:2004qd}. Although these formulations are based on a twisted supersymmetry different from the one described in this paper, evidence has been presented showing that the continuum limit of this formulation for the
case of two-dimensional SYM with four supercharges \cite{Kanamori:2008bk, Hanada:2009hq} yields the same physics \cite{Hanada:2010qg}.

The lattice action of $\cN = 4$ SYM retains invariance under only the single scalar supercharge $\cQ$. The remaining fifteen supercharges $\cQ_a, \cQ_{ab}$, $a, b = 1, \ldots, 5$, are not exact symmetries of the lattice action but are broken by terms of order the lattice spacing. Thus one must potentially tune a series of lattice couplings (counterterms) to achieve full supersymmetry in the continuum limit. The supersymmetric Ward-Takahashi (WT) identities corresponding to $\cQ_a, \cQ_{ab}$ would seem to be very important in this regard as they would provide insight into the degree of breaking and would serve as indicators that the full SUSY has been restored, during the process of fine-tuning. However, even in the continuum theory there are issues with these WT identities. There is no fully off-shell formulation of $\cN = 4$ SYM, so when one goes through the normal process of deriving the WT identities, an essential ingredient is missing: the action is not invariant under the global SUSY transformation, unless the equations of motion are used. However, in the quantum theory (and in particular under the path integral) the classical equations of motion are not satisfied. This casts a pall over the whole procedure, and actually raises questions about whether or not $\langle 0 | \partial_\mu S^\mu_A(x) {\cal O}(0) | 0 \rangle$ really vanishes for $x \not= 0$ and $A = a$ or $ab$, since these symmetries are not realized off-shell\footnote{Here, $S_A^\mu$ is the supercurrent and ${\cal O}$ is any operator.}. Another issue, which arises in the lattice theory, is that because the supercurrents are not conserved at finite lattice spacing (due to discretization errors that break these 15 supersymmetries), the bare supercurrent (i.e., the one simply transcribed from the continuum) will mix with other operators under renormalization. This has been extensively discussed in the context of $\cN = 1$ SYM with Wilson fermions \cite{Farchioni:2001wx} and in another formulation of $\cN = 4$ SYM \cite{Elliott:2008jp}. Thus to use the SUSY WT identities in order to fine-tune the action, one must simultaneously determine these mixing coefficients. In the theory that we are discussing here, there are many such operators and mixing coefficients, rendering this a daunting task.

To our great relief, we have discovered a way to circumvent these difficulties! We have found an intimate relationship between the restoration of the $SU(4)_R$ global R-symmetry of the continuum theory and the supersymmetric continuum limit of the lattice theory. In short, the exact $\cQ$ SUSY, when combined with a restoration of the R-symmetry (or in fact, even a discrete subgroup of that symmetry), is sufficient to guarantee that $\cQ_{a}, \cQ_{ab}$ are restored and the long distance effective action is $\cN = 4$ supersymmetric. So, rather than measuring the SUSY WT identities, one can simply measure the $SU(4)_R$ WT identies. (See Ref. \cite{Kanamori:2008bk} where the authors have constructed and numerically measured WT identities related to the supercurrent conservation in the two-dimensional four supercharge SYM.) Moreover, because of the twisting process, which combines this internal global symmetry with the spacetime symmetry group of Euclidean $SO(4)$ rotations, we ultimately find the remarkable result that the recovery of Lorentz invariance in the continuum limit likely implies the full $\cN = 4$ supersymmetry without any additional fine-tuning. However, it is an open question as to whether or not the $SO(4)$ rotational symmetry requires fine-tuning of the lattice action in the present case. We show what condition must hold on coefficients in the long distance effective theory for this to be the case. Even if fine-tuning of these coefficients is necessary, at least it is relatively easy in a lattice simulation to test for $SO(4)$ invariance at long distance.

The outline of the paper is as follows. We begin with a discussion of the continuum twisted $\cN = 4$ SYM theory and its supersymmetries in Sec.~\ref{sec:4d-sym}. The action of the theory and the scalar supersymmetry transformations of the fields are written down. We then derive the form of the fifteen additional twisted supersymmetries by combining a set of discrete R-symmetries of the continuum theory with the action of the scalar supersymmetry. In Sec.~\ref{sec:R_symm-tune-action} we discuss how invariance of the action under these discrete R-symmetries is already sufficient to guarantee invariance under the remaining fifteen twisted supersymmetries. In Sec.~\ref{sec:wti-sc} we briefly detail the difficulty with the Ward-Takahashi identities for the supercurrent in the twisted theory, due to the fact that there is no fully off-shell formulation\footnote{There is a similar problem in the untwisted theory, again due to the lack of an off-shell formulation.}. We then discuss the lattice discretization of the theory in Sec.~\ref{sec:susy_lattice}. We discuss the fate of the discrete R-symmetries on the lattice in Sec.~\ref{sec:Rsym-and-latt} and show that while they are not invariances of the lattice theory they may be restored {\it automatically} in an appropriate rotationally symmetric continuum limit. Finally, we provide a summary of our main results in Sec.~\ref{sec:conclusions}.

%###############################################################
\section{The continuum $\cN = 4$ theory and its supersymmetries}
\label{sec:4d-sym}
%###############################################################

The four-dimensional Euclidean $\cN = 4$ Yang-Mills theory in the continuum is obtained by dimensional reduction of ten-dimensional $\cN = 1$ SYM. The ten-dimensional theory has a rotation group $SO(10)_{\rm Lorentz}$ and in the four-dimensional theory it becomes the product of the four-dimensional rotation group and the R-symmetry group: $SO(4)_{\rm Lorentz} \times~SO(6)_{\rm R}$. The four-dimensional theory can be rewritten in a useful way by transforming to a new set of twisted fields as shown originally by Marcus \cite{Marcus:1995mq}\footnote{This twist of $\cN = 4$ SYM also plays a role in the geometric-Langlands program \cite{Kapustin:2006pk}.}. The action of the four-dimensional theory, after the twist, can be written in a compact form that can be seen as the dimensional reduction of a five-dimensional theory (see \cite{Catterall:2007kn} for details of the construction):
\bea
\label{eq:compact-five-d-action}
S &=& \int \Tr \Big( -\cFb_{mn} \cF_{mn} + \hf [\cDb_m, \cD_m]^2 - \chi_{mn} \cD_{[m}\psi_{n]} - \eta\cDb_m \psi_m \nn \\
&&- \qtr\epsilon_{mnrpq} \chi_{pq} \cDb_r \chi_{mn}\Big),
\eea
where the Roman indices run from $1, \cdots, 5$. Here and elsewhere below there is an implicit $d^4x$ integration measure. The bosonic degrees of freedom of the theory, the four-dimensional gauge field and six scalars, are encoded in the five complexified gauge fields $\cA_m$, with $\cA_m = A_m + iB_m$. This encoding of fields leads to complexified covariant derivatives and field strengths in the theory. We have, for a generic field $X$  
\beq
\cD_m X \equiv \partial_m X + [\cA_m, X~],~~\cDb_m X \equiv \partial_m X+ [\cAb_m, X~],
\eeq
and the complexified field strengths
\bea
\cF_{mn} &\equiv& [\cD_m, \cD_n] = \partial_m \cA_n - \partial_n \cA_m + [\cA_m, \cA_n], \\
\cFb_{mn} &\equiv& [\cDb_m, \cDb_n] = \partial_m \cAb_n - \partial_n \cAb_m + [\cAb_m, \cAb_n].
\eea
Of course, by definition $\partial_5 = 0$ since the theory has been reduced to four dimensions. The fermions of the original theory become p-forms after twisting. The 16 fermionic degrees of freedom are now packaged in the set ($\eta/2$, $\psi_m$, $\chi_{mn}$) of p-form fields.

One of the most important aspects of the twisting process is the exposure of a nilpotent scalar supercharge $\cQ$ that would make the theory compatible with lattice discretization. The remaining supersymmetric charges also transform as integer spin representations of the twisted rotation group. They are packaged as the p-forms ($\cQ$, $\cQ_m$, $\cQ_{mn}$).

The nilpotent scalar supersymmetry $\cQ$ acts on the twisted fields in a simple manner
\bea
\label{eq:aa-psia}
\cQ \cA_m &=& \psi_m, \\
\cQ \psi_m &=& 0, \\
\cQ \cAb_m &=& 0, \\
\cQ \chi_{mn} &=& -[\cDb_m, \cDb_n], \\
\cQ \eta &=& d, \\
\label{eq:dzero}
\cQ d &=& 0,
\eea
where $d$ is an auxiliary field introduced for the off-shell completion of the 
action.\footnote{It should be emphasized that the action is only off-shell with
respect to $\cQ$, whereas the other 15 supersymmetries require the
equations of motion to yield invariance.}
It obeys the equation of motion 
\beq
d = \sum_m [\cDb_m, \cD_m].
\eeq

The off-shell form of the action can be written as the sum of a $\cQ$-exact piece and a $\cQ$-closed piece. It is
\beq
S = \int \Tr \cQ \Lambda + S_{\cQ-\rm closed},
\eeq
where
\beq
\Lambda = \chi_{mn} [\cD_m, \cD_n] + \eta [\cDb_m, \cD_m] - \hf \eta d,
\eeq
and
\beq
S_{\cQ-\rm closed} = - \qtr \int \Tr \epsilon_{mnrpq} \chi_{pq} \cDb_r \chi_{mn}.
\eeq
We recover the on-shell form of the action given in Eq. (\ref{eq:compact-five-d-action}) after integrating out the auxiliary field $d$. The off-shell action is invariant under the $\cQ$ supersymmetry 
\beq
\cQ S = 0,
\eeq
as the $\cQ$-exact term in the off-shell action vanishes due to the nilpotent property of $\cQ$ and the $\cQ$-closed term is zero due to Bianchi identity for the complexified covariant derivatives.

The twisted action Eq. (\ref{eq:compact-five-d-action}) possesses global discrete symmetries that are subgroups of the R-symmetry group $SO(6)_{\rm R}$ of the theory. We can make use of these discrete symmetries and the $\cQ$ supercharge transformations of the fields Eq. (\ref{eq:aa-psia}) - (\ref{eq:dzero}) to write down the field transformations associated with the $\cQ_a$ and $\cQ_{ab}$ supersymmetries.

%##############################################################
\subsection{Action of the R-symmetries on the twisted fields}
\label{subsec:R_sym}
%##############################################################

We will show that the twisted supersymmetries can be recovered from a knowledge of the scalar supersymmetry $\cQ$ and a discrete subgroup of the R-symmetries. To this end let us first show how to find these discrete R-symmetries. Consider first the (kinetic part of) twisted continuum fermion action written in matrix form
\beq
S_F = \int \Tr \left( \overline{\Psi} \gamma_m \partial_m \Psi \right).
\eeq
Here, $\Psi$ is the $4\times 4$ matrix that results from twisting the $SO(4)$ rotational symmetry with a $SO(4)$ subgroup of the $SO(6)$ R-symmetry. Consider the following transformation on this matrix fermion
\beq
\Psi \to \Psi \Gamma^A,
\eeq 
where $\Gamma^A$ with $A = 1,\cdots, 15,$ belong to the set $\{i\gamma_m,\hf [\gamma_m,\gamma_n]\}$ of $4 \times 4$ matrices that form a basis for the generators of the $SU(4) \simeq SO(6)$ R-symmetry. Furthermore, since the fermions are Majorana $\overline{\Psi} = C \Psi^T C^{-1}$ one can easily show that $\overline{\Psi} \to -\Gamma^A \overline{\Psi}$ and hence that the fermion action is invariant under the transformation. Expanding the matrix $\Psi$ on the same basis (plus the unit matrix) yields the usual twisted fields $(\eta/2, \psi_m,\chi_{mn})$. It should be clear that these twisted fields are then permuted in a well defined manner after right multiplication by one of the $\Gamma^A$. If one subsequently demands that the Yukawa interactions are also invariant one finds that the complexified covariant derivatives must also transform according to the rule:
\beq
{\rm Im}\; \cD_m \to \qtr {\rm Tr} \left( -\Gamma^A \gamma_m \Gamma^A\right ){\rm Im}\; \cD_m,
\eeq
where ${\rm Im}\; \cD_m \equiv (\cD_m-\cDb_m)/2i$. (See also \cite{D'Adda:2005zk} where the authors have shown that a solution of the above consistency condition exactly describes which charge can be defined on a lattice site.) In the next sections we will write out these discrete transformations explicitly and verify the invariance of the continuum action.

%###########################################################
\subsection{$\cQ_a$ supersymmetries}
\label{subsec:Qa_symm}
%###########################################################

Consider the following set of transformations of the fields and covariant derivatives corresponding to the discrete R-transformation associated with $i\gamma_a$ (the index $a$ is considered fixed in all that follows)
\bea
\label{eq:eta-psia}
\eta &\rightarrow& 2 \psi_a, \\
\psi_a &\rightarrow& \hf \eta, \\
\psi_b &\rightarrow& -\chi_{ab}, \\
\chi_{ab} &\rightarrow& -\psi_{b}, \\
\chi_{bc} &\rightarrow& \hf \epsilon_{bcagh}\chi_{gh},
\eea
and
\bea
\cD_a &\rightarrow& \cD_a,~~\cDb_a \rightarrow \cDb_a, \\
\label{eq:db-dbbar}
\cD_b &\rightarrow& \cDb_b,~~\cDb_b \rightarrow \cD_b,~~b \neq a.
\eea

We denote this set of discrete transformations as $R_a$ transformations. Let us check whether the action is invariant under $R_a$ transformations. We have the bosonic action
\bea
S_B &=& \int \Tr -\sum_{m,n} \cFb_{mn} \cF_{mn} + \hf \Big(\sum_m [\cDb_m, \cD_m]\Big)^2.
\eea

Rewriting this to expose the index $a$,
\bea
S_B &=& \int \Tr -2\sum_{n \neq a}[\cDb_a, \cDb_n] [\cD_a, \cD_n] -\sum_{m \neq a,n \neq a} [\cDb_m, \cDb_n] [\cD_m, \cD_n] \nn \\
&&+ \hf \Big([\cDb_a, \cD_a] + \sum_{m \neq a} [\cDb_m, \cD_m]\Big)^2.
\eea

Applying the above field transformations the bosonic action becomes
\bea
S^\prime_B &=& \int \Tr -2\sum_{n \neq a}[\cDb_a, \cD_n] [\cD_a, \cDb_n] - \sum_{\{m, n\} \neq a} [\cD_m, \cD_n] [\cDb_m, \cDb_n] \nn \\
&&+ \hf \Big([\cDb_a, \cD_a] - \sum_{m \neq a} [\cDb_m, \cD_m]\Big)^2 \nn \\
&=& \int \Tr -2 \sum_{n \neq a}[\cDb_a, \cD_n] [\cD_a, \cDb_n] - \sum_{\{m, n\} \neq a} [\cD_m, \cD_n] [\cDb_m, \cDb_n] + 
\hf [\cDb_a, \cD_a]^2 \nn \\
&&+ \hf \sum_{m \neq a} [\cDb_m, \cD_m]^2 - \sum_{m \neq a}[\cDb_a, \cD_a][\cDb_m, \cD_m].
\eea

Thus the change in the bosonic action under this discrete transformation is
\bea
S^\prime_B - S_B &=& \int \sum_{n \neq a} \Tr \Big( -2[\cDb_a, \cD_n] [\cD_a, \cDb_n] + 2[\cDb_a, \cDb_n] [\cD_a, \cD_n] \nn \\
&&- 2[\cDb_a, \cD_a][\cDb_n, \cD_n] \Big).
\eea

Making use of the following relation among complexified covariant derivatives under the trace
\beq
\Tr \Big([\cDb_a, \cD_n][\cD_a, \cDb_n]\Big) = \Tr \Big([\cDb_a, \cDb_n][\cD_a, \cD_n]\Big) - \Tr \Big([\cDb_a, \cD_a][\cDb_n, \cD_n]\Big),
\eeq
with $n\neq a$, it is straightforward to show that $S^\prime_B-S_B = 0$, i.e., the bosonic action is indeed invariant under this discrete transformation.

Let us check the invariance of the fermionic action. We have
\bea
S_F &=& \int \Tr - \eta \sum_m \cDb_m \psi_m - \sum_{m, n} \chi_{mn} \cD_{[m}\psi_{n]} - \qtr \sum_{m,n,c,d,e} \epsilon_{mncde} \chi_{de} \cDb_c \chi_{mn}.
\eea

Exposing the index $a$, 
\bea
S_F &=& \int \Tr \Big(-\eta \cDb_a \psi_a - \sum_{n \neq a} \eta \cDb_n \psi_n -2 \sum_{n \neq a} \chi_{an} \cD_a\psi_n + 2\sum_{n \neq a} \chi_{an} \cD_n\psi_a \nn \\
&&-\sum_{\{m, n \} \neq a} \chi_{mn} \cD_m\psi_n + \sum_{\{m, n \} \neq a} \chi_{mn} \cD_n\psi_m \nn \\
&&- \sum_{\{n,c,d,e\} \neq a} \epsilon_{ancde} \chi_{de} \cDb_c \chi_{an} - \qtr \sum_{\{m,n,d,e\} \neq a} \epsilon_{mnade} \chi_{de} \cDb_a \chi_{mn}\Big).
\eea

Under the $R_a$ transformations it becomes
\bea
S'_F &=& \int \Tr \Big(-2 \psi_a \cDb_a \hf \eta + \sum_{n \neq a} 2 \psi_a \cD_n \chi_{an} - 2 \sum_{n \neq a} \psi_n \cD_a\chi_{an} - 2 \sum_{n \neq a} \psi_n \cDb_n \hf \eta \nn \\
&&+\sum_{\{m, n\} \neq a} \hf \epsilon_{mnagh}\chi_{gh} \cDb_m\chi_{an} - \sum_{\{m, n\} \neq a} \hf \epsilon_{mnagh}\chi_{gh} \cDb_n\chi_{am} \nn \\
&&+ \epsilon_{ancde} \hf \epsilon_{deapq} \chi_{pq} \cD_c \psi_n - \qtr \epsilon_{mnade} \hf \epsilon_{deapq}\chi_{pq} \cDb_a \hf \epsilon_{mnars}\chi_{rs} \Big) \nn \\
&=& \int \Tr \Big(-\eta \cDb_a \psi_a + \sum_{n \neq a} 2 \chi_{an} \cD_n \psi_a - \sum_{n \neq a} 2 \chi_{an} \cD_a \psi_n - \sum_{n \neq a}\eta \cDb_n \psi_n \nn \\
&&+\sum_{\{m, n\} \neq a} \epsilon_{mnagh}\chi_{gh} \cDb_m\chi_{an} - \sum_{\{n, c\} \neq a}\chi_{nc} \cD_n \psi_c + \sum_{\{n, c\} \neq a} \chi_{nc} \cD_c \psi_n \nn \\
&&- \qtr \epsilon_{pqars} \chi_{pq} \cDb_a \chi_{rs} \Big) \nn \\
&=& S_F.
\eea
Thus the fermionic action is also invariant under the $R_a$ transformations.

Combining this discrete R-symmetry with the original $\cQ$ supersymmetry we can now write down the one-form supersymmetries associated with the twisted supercharge $\cQ_a$. After making the replacements Eq. (\ref{eq:eta-psia}) - (\ref{eq:db-dbbar}) in Eq. (\ref{eq:aa-psia}) - (\ref{eq:dzero}) we have the supersymmetry transformations associated with the supercharge $\cQ_a$:
\bea
\cQ_a \cA_b &=& \hf \delta_{ab} \eta, \\
\cQ_a \cAb_b &=& -\chi_{ab}, \\
\cQ_a \psi_b &=& \hf \delta_{ab} d_a + (1-\delta_{ab})[\cDb_a, \cD_b], \\
\cQ_a \chi_{bc} &=& -\hf \epsilon_{abcgh}[\cD_g, \cD_h], \\
\cQ_a \eta &=& 0, \\
\cQ_a d_a &=& 0.
\eea
Here, $d_a = [ \cDb_a , \cD_a ] - \sum_{m \not= a} [ \cDb_m , \cD_m ]$, and the
equation $\cQ_a d_a = 0$ requires the equations of motion, since this is
an on-shell supersymmetry.

From the above supersymmetry transformations we see that the $\cQ_a$, like the scalar supercharge $\cQ$, is also nilpotent: $\cQ_a^2 = 0$. A straightforward, if tedious, calculation allows one to verify that the action is $\cQ_a$ invariant on-shell (i.e., using the equations of motion), $\cQ_a S = 0$.

%#############################################################
\subsection{$\cQ_{ab}$ supersymmetries}
\label{subsec:Qab_sym}
%#############################################################

We now consider the discrete subgroup of the R-symmetries associated with $\Gamma = \hf [\gamma_a, \gamma_b]$ corresponding to transformations of the twisted fields and covariant derivatives parametrized by the fixed indices $a$ and $b$. 
\bea
\label{eq:eta-chiab}
\eta &\rightarrow& 2 \chi_{ab}, \\
\psi_a &\rightarrow& \psi_b, \\
\psi_b &\rightarrow& -\psi_a, \\
\psi_c &\rightarrow& \hf \epsilon_{cabgh}\chi_{gh}, \\
\chi_{ab} &\rightarrow& -\hf \eta, \\
\chi_{ac} &\rightarrow& \chi_{bc}, \\
\chi_{bc} &\rightarrow& -\chi_{ac}, \\
\chi_{gh} &\rightarrow& - \epsilon_{ghabc}\psi_c,
\eea
and
\bea
\cD_{a,b} &\rightarrow& \cDb_{a,b},~~\cDb_{a,b} \rightarrow \cD_{a,b}, \\
\label{eq:dbc-dbc}
\cD_c &\rightarrow& \cD_c,~~\cDb_c \rightarrow \cDb_c,~~c \neq a,b.
\eea

We denote this set of discrete R-symmetries as $R_{ab}$ transformations. Let us check if $R_{ab}$ transformations are indeed a symmetry of the action. First consider the bosonic action
\bea
S_B &=& \int \Tr -\sum_{m,n} \cFb_{mn} \cF_{mn} + \hf \Big(\sum_m [\cDb_m, \cD_m]\Big)^2.
\eea

Splitting this into pieces to expose the terms with indices $a$ and $b$
\bea
S_B &=& \int \Tr \Big( -2[\cDb_a, \cDb_b] [\cD_a, \cD_b] - 2 \sum_{c \neq a, b} [\cDb_a, \cDb_c] [\cD_a, \cD_c] - 2 \sum_{c \neq a, b} [\cDb_b, \cDb_c] [\cD_b, \cD_c] \nn \\
&&- \sum_{\{c, d\} \neq a,b} [\cDb_c, \cDb_d] [\cD_c, \cD_d] + \hf \Big([\cDb_a, \cD_a] + [\cDb_b, \cD_b] + \sum_{c \neq a, b}[\cDb_c, \cD_c] \Big)^2 \Big).
\eea

Applying the field transformations
\bea
S^{\prime \prime}_B &=& \int \Tr \Big( -2[\cD_a, \cD_b] [\cDb_a, \cDb_b] - 2\sum_{c \neq a, b} [\cD_a, \cDb_c] [\cDb_a, \cD_c] - 2\sum_{c \neq a, b}[\cD_b, \cDb_c] [\cDb_b, \cD_c] \nn \\
&&- \sum_{\{c,d\} \neq a,b} [\cDb_c, \cDb_d] [\cD_c, \cD_d] + \hf \Big(-[\cDb_a, \cD_a] - [\cDb_b, \cD_b] + \sum_{c \neq a, b}[\cDb_c, \cD_c] \Big)^2 \Big).
\eea

Thus the change in the bosonic action under $R_{ab}$ transformations is
\bea
S^{\prime \prime}_B - S_B &=& \int \Tr \Big( - 2\sum_{c \neq a, b} [\cD_a, \cDb_c] [\cDb_a, \cD_c] - 2\sum_{c \neq a, b}[\cD_b, \cDb_c] [\cDb_b, \cD_c] \nn \\
&&+ 2\sum_{c \neq a, b} [\cDb_a, \cDb_c] [\cD_a, \cD_c] + 2\sum_{c \neq a, b}[\cDb_b, \cDb_c] [\cD_b, \cD_c] \nn \\
&&- 2\sum_{c \neq a, b}[\cDb_a, \cD_a][\cDb_c, \cD_c] - 2\sum_{c \neq a, b}[\cDb_b, \cD_b][\cDb_c, \cD_c] \Big).
\eea

Once again the relation
\beq
\Tr ([\cDb_a, \cD_b][\cD_a, \cDb_b]) = \Tr ([\cDb_a, \cDb_b][\cD_a, \cD_b]) - \Tr ([\cDb_a, \cD_a][\cDb_b, \cD_b]),
\eeq
allows us to show that in fact $S^{\prime\prime}_B - S_B = 0$ and thus the bosonic action is invariant under this discrete field transformation.

Turning to the fermionic part of the action we have
\bea
S_F &=& \int \Tr \Big(-\eta \sum_m \cDb_m \psi_m -\sum_{m, n} \chi_{mn} \cD_{[m}\psi_{n]} - \qtr \sum_{m,n,c,d,e} \epsilon_{mncde} \chi_{de} \cDb_c \chi_{mn} \Big).
\eea

Splitting this into pieces to expose the terms with indices $a$ and $b$
\bea
S_F &=& \int \Tr \Big(-\eta \cDb_a\psi_a -\eta \cDb_b\psi_b -\sum_{m \neq a, b} \eta \cDb_m\psi_m - 2\chi_{ab} \cD_a\psi_b + 2\chi_{ab} \cD_b\psi_a \nn \\
&&- 2\sum_{n \neq a, b} \chi_{an} \cD_a\psi_n + 2\sum_{n \neq a, b} \chi_{an} \cD_n\psi_a - 2 \sum_{n \neq a, b} \chi_{bn} \cD_b\psi_n + 2\sum_{n \neq a, b} \chi_{bn} \cD_n\psi_b \nn \\
&&- \sum_{\{m, n\} \neq a, b}\chi_{mn} \cD_m\psi_n + \sum_{\{m, n\} \neq a, b}\chi_{mn} \cD_n\psi_m - \epsilon_{abcde} \chi_{ab} \cDb_c \chi_{de} \nn \\
&&- 2 \epsilon_{ancbe} \chi_{an} \cDb_c \chi_{be} - \epsilon_{deabn} \chi_{de} \cDb_a \chi_{bn} - \epsilon_{deban} \chi_{de} \cDb_b \chi_{an} \Big).
\eea

Applying the field transformations given in Eq. (\ref{eq:eta-chiab}) - (\ref{eq:dbc-dbc}) the fermionic action becomes
\bea
S^{\prime\prime}_F &=& \int \Tr \Big(- 2 \chi_{ab} \cD_a\psi_b + 2 \chi_{ab} \cD_b\psi_a - \chi_{ab} \cDb_m\epsilon_{mabgh}\chi_{gh} - \eta \cDb_a\psi_a - \eta \cDb_b\psi_b \nn \\
&&- \chi_{bn} \cDb_a\epsilon_{nabgh}\chi_{gh} + 2 \chi_{bn} \cD_n\psi_b + \chi_{an} \cDb_b\epsilon_{nabgh}\chi_{gh} + 2 \chi_{an} \cD_n\psi_a \nn \\
&&+ \hf \epsilon_{mnabc} \psi_c \cD_m\epsilon_{nabrs}\chi_{rs} - \hf \epsilon_{mnabh}\psi_h \cD_n\epsilon_{mabuv}\chi_{uv} - \hf \epsilon_{abcde} \eta \cDb_c\epsilon_{deabh}\psi_h \nn \\
&&+ 2 \epsilon_{ancbe} \chi_{bn} \cDb_c\chi_{ae} - \epsilon_{deabn} \epsilon_{deabc}\psi_c \cD_a\chi_{an} + \epsilon_{deban} \epsilon_{deabp}\psi_p \cD_b\chi_{bn} \Big).
\eea

Simplifying and rearranging the terms
\bea
S^{\prime\prime}_F &=& \int \Tr \Big(- \eta \cDb_a\psi_a - \eta \cDb_b\psi_b - 2 \eta \cDb_m\psi_m - 2 \chi_{ab} \cD_a\psi_b + 2 \chi_{ab} \cD_b\psi_a \nn \\
&&- 2 \chi_{an} \cD_a\psi_n + 2 \chi_{an} \cD_n\psi_a - 2 \chi_{bn} \cD_b\psi_n + 2 \chi_{bn} \cD_n\psi_b \nn \\
&&- \chi_{mn} \cD_m \psi_n + \chi_{mn} \cD_n\psi_m - \epsilon_{abcde} \chi_{ab} \cDb_c\chi_{de} \nn \\
&&- 2 \epsilon_{ancbe} \chi_{an} \cDb_c\chi_{be} - \epsilon_{deabn} \chi_{de} \cDb_a\chi_{bn} - \epsilon_{deban} \chi_{de} \cDb_b\chi_{an} \Big)\nn \\
&=& S_F.
\eea

Thus we have recovered the original fermionic part of the action.

After making the replacements Eq. (\ref{eq:eta-chiab}) - (\ref{eq:dbc-dbc}) in the $\cQ$ supersymmetries for the fields we have the supersymmetry transformations associated with the supercharge $\cQ_{ab}$:
\bea
\cQ_{ab} \cA_c &=& \hf \sum_{g, h} \epsilon_{abcgh}\chi_{gh}, \\
\cQ_{ab} \cAb_c &=& \delta_{ac}\psi_b - \delta_{bc}\psi_a, \\
\cQ_{ab} \psi_c &=& \sum_{g, h} \epsilon_{abcgh}[\cDb_g, \cDb_h], \\
\cQ_{ab} \chi_{cd} &=& \hf \delta_{ac}\delta_{bd} d_{ab} 
- \delta_{bc}[\cD_a, \cDb_d] + \delta_{ac}[\cD_b, \cDb_d], \\
\cQ_{ab} \eta &=& 2[\cD_a, \cD_b], \\
\cQ_{ab} d_{ab} &=& 0.
\eea
Here, $d_{ab} = -[ \cDb_a , \cD_a ] - [ \cDb_b , \cD_b ] + \sum_{m \not= a,b} [ \cDb_m , \cD_m ]$, and the
equation $\cQ_{ab} d_{ab} = 0$ requires the equations of motion, since this is
an on-shell supersymmetry.

From the above supersymmetry transformations we see that the supercharge $\cQ_{ab}$ is also nilpotent: $\cQ_{ab}^2 = 0$ and the action invariant under $\cQ_{ab}$ on-shell. 

%####################################################################
\subsection{Combined expressions for supersymmetry transformations}
\label{subsec:combined-susy}
%####################################################################

The supersymmetry transformations can be combined to express them in a compact form once we introduce the (Grassmann odd) supersymmetry parameters $(\delta\kappa_0, \delta\kappa_a, \delta\kappa_{ab})$ associated with the supercharges $(\cQ, \cQ_a, \cQ_{ab})$. Thus the supersymmetry transformations of the fields are
\bea
\label{eq:transforms1}
\delta_\kappa \cA_m &=& \delta \kappa_0 \psi_m + \hf \delta \kappa_b \delta_{bm}\eta + \hf \delta \kappa_{ab} \epsilon_{abmgh}\chi_{gh}, \\
\delta_\kappa \cAb_m &=& -\delta \kappa_b \chi_{bm} + \delta \kappa_{ab}(\delta_{am}\psi_b - \delta_{bm}\psi_a), \\
\delta_\kappa \psi_m &=& \delta \kappa_b \Big(\hf \delta_{bm}d_b + (1-\delta_{bm})[\cDb_b, \cD_m]\Big) + \delta \kappa_{ab} \epsilon_{abmgh}[\cDb_g, \cDb_h], \\
\delta_\kappa \chi_{mn} &=& -\delta \kappa_0 [\cDb_m, \cDb_n] - \hf \delta \kappa_b \epsilon_{bmngh}[\cD_g, \cD_h] \nn \\
&&+ \delta \kappa_{ab}\Big(\hf \delta_{am}\delta_{bn} d_{ab} - \delta_{bm}[\cD_a, \cDb_n] + \delta_{am}[\cD_b, \cDb_n]\Big),\\
\delta_\kappa \eta &=& \delta \kappa_0 d + 2\delta \kappa_{ab} [\cD_a, \cD_b], \\
\label{eq:transforms2}
\delta_\kappa d &=& \delta_\kappa d_a = \delta_\kappa d_{ab} = 0
\eea
where the operator $\delta_\kappa$ is
\beq
\delta_\kappa = \delta \kappa_0 \cQ + \delta \kappa_a \cQ_a + \delta \kappa_{ab}\cQ_{ab}.
\eeq

The above set of transformations are clearly a symmetry of the action if the parameters $(\delta\kappa_0, \delta\kappa_a, \delta\kappa_{ab})$ are strictly singlets under the gauge group. 

%######################################################
\section{R-symmetries and fine-tuning of the action}
\label{sec:R_symm-tune-action}
%######################################################

In Ref. \cite{Catterall:2011pd} it has been shown that the only relevant counterterms that can modify the action through radiative corrections correspond to renormalizations of marginal operators already present in the bare action. Thus the most general form for the renormalized action is
\bea
S &=& \int \Tr \cQ \Big(\alpha_1 \chi_{mn}\cF_{mn} + \alpha_2 \eta [\cDb_m, \cD_m] 
- \hf \alpha_3  \eta d \Big) - \qtr \alpha_4 \epsilon_{mncde}\chi_{de}\cDb_c\chi_{mn},
\eea
where $\alpha_i$ with $i = 1, \cdots, 4$ are dimensionless numbers taking values $(1, 1, 1, 1)$ in the classical theory. Indeed one would think that at most three dimensionless 
ratios of these couplings might need to be tuned in the quantum theory. In this section we show that the discrete R-symmetries provide a powerful constraint on the relations between $\alpha_i$'s since the operators appear through radiative corrections to the action must respect these additional symmetries.

After integrating out the auxiliary field $d$, we have the form of the action with general coefficients
\bea
S &=& \int\Tr \Big( -\alpha_1 \cFb_{mn} \cF_{mn} + \hf \Big(\frac{\alpha_2^2}{\alpha_3}\Big)[\cDb_m, \cD_m]^2 - \alpha_1\chi_{mn} \cD_{[m}\psi_{n]} - \alpha_2\eta\cDb_m \psi_m \nn \\
&&- \qtr\alpha_4\epsilon_{mnrpq} \chi_{pq} \cDb_r \chi_{mn}\Big).
\label{ldefa}
\eea

We have shown that the action is invariant under the $R_a$ transformations. The invariance under
any one of the $R_a$ transformations in the bosonic sector yields the following constraint
\beq
\alpha_1 = \frac{\alpha_2^2}{\alpha_3}.
\eeq
In the fermionic sector invariance under any one of the $R_a$ transformation gives another constraint
\beq
\alpha_1 = \alpha_2 = \alpha_4.
\eeq
Combining these two constraints we find
\beq
\label{eq:alphai_alpha}
\alpha_i = \alpha,~i=1, \cdots 4,
\eeq
with $\alpha$ an arbitrary parameter, which can be absorbed by redefining the overall gauge coupling of the theory.

A similar set of relations among $\alpha_i$ and the same 
conclusion as in Eq.~(\ref{eq:alphai_alpha}) can be obtained 
from examining the invariance of the action under any one of 
the $R_{ab}$ interchange symmetries. 

Thus we can write the quantum action of the theory as
\bea
S &=& \int\Tr \alpha \Big( -\cFb_{mn} \cF_{mn} + \hf [\cDb_m, \cD_m]^2 - \chi_{mn} \cD_{[m}\psi_{n]} - \eta\cDb_m \psi_m \nn \\
&&- \qtr \epsilon_{mnrpq} \chi_{pq} \cDb_r \chi_{mn}\Big).
\eea
These results imply that invariance under the single scalar 
supercharge $\cQ$ together with invariance under any one of the 15 $R_a$ or $R_{ab}$ 
symmetries implies that the theory is invariant under {\it all} additional 
twisted supersymmetries.  We will use this fact to argue later that the 
supersymmetries of the lattice theory will be enhanced at any point 
in the parameter space where the R-symmetries of the theory are restored.

It is interesting to note that if $SO(4)$ rotational
symmetry (the Euclidean version of Lorentz invariance)
is restored in the continuum limit, then the supersymmetry
must be enhanced to at least $\cN=1$.  In that case it is quite possible
that a discrete R-symmetry will be present since $\cN=1$ theories
often have a $U(1)_R$ R-symmetry---and even if it is anomalous
a discrete subgroup survives.  We would then expect to have one or more of
the 15 $R_a$ or $R_{ab}$ symmetries in effect.  In that case the full
$\cN=4$ supersymmetry emerges because of the above arguments.

%################################################################
\section{Ward-Takahashi identities}
\label{sec:wti-sc}
%################################################################

In this section we illustrate the difficulty that occurs when we
attempt to derive the WT identities involving the supercurrent and any 
local (or non-local) operator by making a change of variables in the 
functional integral which defines the expectation value of the operator. 
We will find that the fact that we do not have a fully off-shell formulation
presents an obstruction.

Let us consider a composite operator $O(y)$ constructed out of the 
component fields $\Phi$ of the theory. The expectation value of this operator is given by
\beq
\langle O(y) \rangle = \frac{1}{\cZ} \int [d\Phi] e^{-S[\Phi]} O(y),
\eeq
where $\cZ$ denotes the partition function and $[d\Phi]$ the measure of the functional integral. Now consider some infinitesimal transformation of the fields defined by 
\beq
\Phi(x) \rightarrow \Phi'(x) = \Phi(x) + \delta_\kappa \Phi(x),
\eeq
where $\delta_\kappa \Phi(x) = \kappa \Delta \Phi(x)$, with $\kappa$ a Grassmann parameter 
and $\Delta$ the deformation on the field.

The functional integral itself is independent of any relabeling of integration variables. This gives,
\bea
0 &=& \langle O'(y) \rangle - \langle O(y) \rangle \nn \\
&=& \delta_\kappa \langle O(y) \rangle \nn \\
&=&\frac{1}{\cZ} \int [d\Phi] \Big(e^{-S[\Phi]} (-\delta_\kappa S[\Phi]) O(y) + e^{-S[\Phi]} \delta_\kappa O(y)\Big) \nn \\
&=&- \langle (\delta_\kappa S) O(y) \rangle + \langle \delta_\kappa O(y) \rangle.
\eea
That is,
\beq
\label{eq:WT-relation}
\langle (\delta_\kappa S) O(y) \rangle = \langle \delta_\kappa O(y) \rangle. 
\eeq
For the symmetry transformation $\delta_\kappa  = \kappa_A \cQ_A$, 
with the index $A$ labeling the 16 supercharges, it takes the form
\beq
\langle (\cQ_A S) O(y) \rangle = \langle \cQ_A O(y) \rangle. 
\eeq
So far, the identity presents no problem.  However, in the continuum
one would like to argue that $\cQ_A S = 0$, and so one obtains
the ``global'' WT identity, $\langle \cQ_A O(y) \rangle=0$.  However,
$\cQ_A S = 0$ requires the use of the equations of motion.  These are
not valid for the fields under the path integral, since we are
not just summing over classical paths.  So we have no right to
set $\cQ_A S = 0$, and so $\langle \cQ_A O(y) \rangle=0$ does
not follow.  An exception is the scalar supercharge $\cQ$:  we
do have an off-shell formulation, involving $d$, for which
$\cQ S = 0$ without using the equations of motion.  Thus we find
that we can prove the identity $\langle \cQ O(y) \rangle=0$.

Another comment is in order here.  While it is true that in the
quantum theory we cannot use the classical equations of motion,
we can use their quantum analogue, the Schwinger-Dyson equations.
Then the equations of motion are satisfied up to contact terms.
However $\cQ_A S$ involves an integral over all spacetime,
and so this will always include the point $y$, where the operator
$O(y)$ sits.  Thus we will pick up a nonzero contribution from
the contact term when we evaluate $\langle (\cQ_A S) O(y) \rangle$.
It can be shown that this contact term is precisely the right-hand
side, $\langle \cQ_A O(y) \rangle$.  Thus we end up with a trivial
identity that contains no new information.

For the purpose of finding the critical values of the parameters
for which fine-tuning is successful, and the SUSY long distance
theory emerges, it is more useful to study the ``local'' WT identities---i.e.,
the ones that involve the supercurrent.  The analogy that one
should make is in finding the critical bare mass (equivalently, the critical $\kappa$)
when using Wilson fermions in lattice QCD.  There one studies
the axial current WT identities, such as $\langle 0 | \partial_\mu J_{5\mu}^a(x) P^b(0) | 0 \rangle$,
where $P^b$ is the pseudoscalar density, and demands that they vanish.
To obtain these sorts of identities, we generalize the transformation 
of the fields given above to be a spacetime dependent transformation:
\beq
\Phi(x) \rightarrow \Phi'(x) = \Phi(x) + \delta_{\kappa(x)} \Phi(x),
\eeq
where $\delta_{\kappa(x)} \Phi(x) = \kappa(x) \Delta \Phi(x)$, with $\kappa(x)$ a 
spacetime dependent, arbitrary Grassman function.  In our twisted theory we have
\beq
\delta_{\kappa(x)} = \kappa_A(x) \cQ_A, 
\eeq
with $A$ denoting the sixteen possibilities corresponding to
$\cQ, \cQ_a, \cQ_{ab}$.

If we make use of the equations of motion, the variation of the action can be
brought into the form
\bea
\delta_\kappa S &=& \int d^4x ~\partial_m \kappa_A (x) \cS_A^m(x),
\label{varact}
\eea
where the index $m$ is summed over and $\cS_A^m(x)$ is the supercurrent (Noether current)
associated with $\cN=4$ SUSY.
We know that this is true because if the $\kappa_A$ are constant, the action
is invariant with the equations of motion.
Note that the supercurrent is decomposed into scalar, one-form and two-form parts 
\beq
\cS_A^m = (\cS^m_0, \cS_a^m, \cS_{ab}^m).
\eeq

We write the WT identity relations given in Eq. (\ref{eq:WT-relation}) as
\bea
\langle (\delta_\kappa S) O(y) \rangle &=& \langle \delta_\kappa O(y) \rangle.
\eea
Plugging in the variation of the action,
\bea
- \Big \langle \int d^4 z \kappa_A(z) \Big(\partial_m \cS_A^m(z)\Big) 
O(y) \Big \rangle &=& \Big \langle \kappa_A(y) \cQ_A O(y) \Big \rangle,\nn \\
- \frac{\delta}{\delta\kappa_A(x)} \Big \langle \int d^4 z \kappa_A(z) 
\Big(\partial_m \cS_A^m(z)\Big) O(y) \Big \rangle &=& \frac{\delta}{\delta\kappa_A(x)}
\Big \langle \kappa_A(y) \cQ_A O(y) \Big \rangle,\nn \\
- \Big \langle \int d^4 z \delta^{(4)}(x-z) \partial_m \cS_A^m(z)O(y) 
\Big \rangle &=& \Big \langle \delta^{(4)}(x-y) \cQ_A O(y)\Big \rangle.
\eea
Thus we would have the supersymmetric WT identities
\beq
\label{eq:WTI-final}
\langle \partial_m \cS_A^m(x) O(y) \rangle = -\delta^{(4)}(x-y) \langle \cQ_A O(y) \rangle.
\eeq
Setting $y=0$ they take the form
\beq
\partial_m \langle \cS_A^m(x) O(0) \rangle = -\delta^{(4)}(x) \langle \cQ_A O(0) \rangle.
\eeq
However, note that we made use of the equations of motion in deriving
(\ref{varact}).  This was necessary in order to eliminate terms
which do not involve $\partial_m \kappa_A$, but rather just $\kappa_A$
without derivatives, since the symmetry is only realized on-shell.
Inside the path integral in which we derived the WT identity, the fields
do not generically lie on classical paths and so this step was not
valid.  Thus the proof has a fatal flaw and it is not at all clear that
the WT identity will be satisfied.  However, as above, $\cQ$ is realized
off-shell, so its corresponding identity will in fact be satisfied.
But this is of no use to us in fine-tuning because $\cQ$ is exact on
the lattice anyhow.

%################################################
\section{Twisted supersymmetries on the lattice}
\label{sec:susy_lattice}
%################################################

%

%############################################################
\subsection{The lattice action}
\label{subsec:sym-lattice-n4d4}
%############################################################

The twisted continuum theory can be discretized in a straightforward manner. The complex continuum gauge fields $\cA_a$, $a=1, \cdots, 5$, are mapped to complexified Wilson gauge links $\cU_a$ living on links $(\bn, \bn + \hatbmu_a)$ of a four-dimensional abstract hypercubic lattice, with $\bn$ denoting the integer valued site on the lattice and $\hatbmu_a$ the basis vector along $a$ direction. Since our theory has five complexified gauge links we need a four-dimensional lattice with five basis vectors. We can take all these basis vectors to be equivalent to ensure that the lattice theory enjoys the maximal symmetry. This requirement imposes a condition on the lattice theory that it must possess an $S_5$ point group symmetry. It is indeed consistent with the continuum symmetry of the theory as the $S_5$ point group symmetry is a discrete subgroup of the $SO(4)$ twisted rotational symmetry. A lattice arrangement in which all the five basis vectors are treated equally is called the $A_4^*$ lattice. 

The five basis vectors of the $A_4^*$ lattice are
\bea
\hatbe_1 &=& \left(\frac{1}{\sqrt{2}}, \frac{1}{\sqrt{6}}, \frac{1}{\sqrt{12}}, \frac{1}{\sqrt{20}} \right), \nn \\
\hatbe_2 &=& \left(-\frac{1}{\sqrt{2}}, \frac{1}{\sqrt{6}}, \frac{1}{\sqrt{12}}, \frac{1}{\sqrt{20}} \right), \nn \\
\hatbe_3 &=& \left(0, -\frac{2}{\sqrt{6}}, \frac{1}{\sqrt{12}}, \frac{1}{\sqrt{20}} \right), \\
\hatbe_4 &=& \left(0, 0, -\frac{3}{\sqrt{12}}, \frac{1}{\sqrt{20}} \right), \nn \\
\hatbe_5 &=& \left(0, 0, 0, -\frac{4}{\sqrt{20}} \right), \nn
\eea
and they are aligned such that they connect the center of a 4-simplex to its five corners. It should be noted that these basis vectors are related to the weights of the defining representation of $SU(5)$. The point group symmetry $S_5$ of the lattice action is the Weyl subgroup of $SU(5)$. 

We begin with placing the fields of the lattice theory on the links of an abstract hypercubic lattice with integer site $\bn$ and basis vectors $\hatbmu_a$:
\bea
\label{eq:mu-vectores}
\hatbmu_1 &=& (1, 0, 0, 0),\nn \\
\hatbmu_2 &=& (0, 1, 0, 0),\nn \\
\hatbmu_3 &=& (0, 0, 1, 0), \\
\hatbmu_4 &=& (0, 0, 0, 1),\nn \\
\hatbmu_5 &=& (-1, -1, -1, -1),\nn
\eea
obeying the constraint $\sum_m \hatbmu_m = 0$. These abstract vectors do not correspond to positions in spacetime of sites and links of the $A_4^*$ lattice. They can be related to the physical location in spacetime using the $A_4^*$ basis vectors $\hatbe_a$. We relate the integer site $\bn$ to the spacetime location ${\bf R}$ in the following way:
\beq
{\bf R} = a \sum_{\nu = 1}^4 (\mu_\nu \cdot \bn)\hatbe_\nu = a \sum_{\nu = 1}^4 n_\nu \hatbe_\nu,
\eeq
where $a$ is the lattice spacing. We can also show that a small lattice displacement of the form $d\bn = \hatbmu_m$ on the abstract lattice corresponds to a spacetime translation by $(a\hatbe_m)$ on the $A_4^*$ lattice:
\beq
d{\bf R} = a \sum_{\nu = 1}^4 (\mu_\nu \cdot d\bn)\hatbe_\nu = a \sum_{\nu = 1}^4 (\hatbmu_\nu \cdot \hatbmu_m)\hatbe_\nu = a \hatbe_m.
\eeq
We place the twisted fields on the links of the hypercubic lattice the following way. The field $\psi_m(\bn)$ is the superpartner of $\cU_m(\bn)$ and thus it must also reside on the same link. The field $\cUb_m(\bn)$ resides on the oppositely oriented link from $\bn \to \bn-\hatbmu_m$. The ten fermions $\chi_{mn}(\bn)$ are then chosen to reside on new fermionic links $\bn+\hatbmu_m+\hatbmu_n\to \bn$ while the singlet fermionic field $\eta(\bn)$ is assigned to the degenerate link consisting of a single site $\bn$.
The nilpotent scalar supersymmetry $\cQ$ acts on the lattice fields in the following way
\bea
\cQ \cU_m(\bn) &=& \psi_m(\bn) \\
\cQ \psi_m(\bn) &=& 0 \\
\cQ \cUb_m(\bn) &=& 0 \\
\cQ \chi_{mn}(\bn) &=& -\Big(\cD_m^{(+)}\cU_n(\bn)\Big)^\dagger = -\cF_{mn}^\dagger(\bn) \\
\cQ \eta(\bn) &=& d(\bn) \\
\cQ d(\bn) &=& 0
\eea
We see that the $\cQ$ supersymmetry transforms a bosonic field of one type to a fermionic field of the same type at the same place on the lattice. The complexified field strength takes the following form on the abstract lattice
\beq
\cF_{mn}(\bn) = \cD_m^{(+)}\cU_n(\bn) = \cU_m(\bn)\cU_n(\bn + \hatbmu_m) - \cU_n(\bn)\cU_m(\bn + \hatbmu_n).
\eeq
In the naive continuum limit it reduces to the continuum (complex) field strength and is automatically antisymmetric in the indices.

The link mappings and orientations of the lattice fields are conveniently summarized by giving their gauge transformation properties on the lattice
\bea
\label{eq:gauge-t}
\eta(\bn) &\rightarrow& G(\bn)\eta(\bn)G^{\dagger}(\bn) \\ 
\psi_m(\bn) &\rightarrow& G(\bn)\psi_m(\bn)G^{\dagger}(\bn + \hatbmu_m) \\ 
\chi_{mn} (\bn) &\rightarrow& G(\bn + \hatbmu_m + \hatbmu_n)\chi_{mn}(\bn)G^{\dagger}(\bn) \\ 
\cU_m(\bn) &\rightarrow& G(\bn)\cU_m(\bn)G^{\dagger}(\bn + \hatbmu_m) \\ 
\cUb_m (\bn) &\rightarrow& G(\bn + \hatbmu_m)\cUb_m(\bn)G^{\dagger}(\bn)
\eea
where for instance $G \in U(N)$.  The action on the lattice takes the form
\beq
S = S_{\cQ-\rm exact} + S_{\cQ-\rm closed}~,
\eeq
where
\bea
S_{\cQ-\textrm{exact}} &=& \sum_{\bn} \Tr \Big(-\cF_{mn}^{\dagger}(\bn) \cF_{mn}(\bn) + \hf\Big(\cDb_m^{(-)}\cU_m(\bn)\Big)^2 \nn \\
&&~~~~~~~~- \chi_{mn}(\bn) \cD^{(+)}_{[m}\psi_{n]}(\bn) - \eta(\bn) \cDb^{(-)}_m\psi_m(\bn) \Big),
\eea
and
\beq
S_{\cQ-\rm closed} = \sum_{\bn, m,n,c,d,e} - \qtr \Tr \epsilon_{mncde} \chi_{de}(\bn + \hatbmu_m + \hatbmu_n + \hatbmu_c) \cDb^{(-)}_{c} \chi_{mn}(\bn + \hatbmu_c).
\eeq
We see that the terms in the action form closed loops on the lattice ensuring gauge invariance.

The covariant forward and backward difference operators appearing in the lattice action are given by
\bea
\cD_m^{(+)} f(\bn) &=& \cU_m(\bn) f(\bn + \hatbmu_m) - f(\bn) \cU_m(\bn), \\
\cD_m^{(+)} f_n(\bn) &=& \cU_m(\bn) f_n(\bn + \hatbmu_m) - f_n(\bn) \cU_m(\bn + \hatbmu_n), \\
\cDb_m^{(-)} f_m(\bn) &=& f_m(\bn)\cUb_m(\bn) - \cUb_m(\bn - \hatbmu_m) f_m(\bn - \hatbmu_m), \\
\cDb_c^{(-)} f_{mn} (\bn) &=& f_{mn}(\bn) \cUb_c(\bn - \hatbmu_c) - \cUb_c(\bn + \hatbmu_m + \hatbmu_n - \hatbmu_c) f_{mn}(\bn - \hatbmu_c).
\eea

It can be shown that the $\cQ$-closed term satisfies an exact Bianchi identity on the lattice just as for the continuum
\beq
\epsilon_{mncde}\cDb^{(-)}_c\cFb_{mn}(\bn + \hatbmu_c) = 0.
\eeq

%###################################################################
\subsection{$\cQ_a$ and $\cQ_{ab}$ supersymmetries on the lattice}
\label{subsec:qa-qab-lattice-n4d4}
%###################################################################

If we try to implement Eq. (\ref{eq:transforms1}) - (\ref{eq:transforms2}) on the lattice, according to the discretization prescription given above we encounter an immediate problem - the lattice fields connected by $\cQ_a$ and $\cQ_{ab}$ are located on distinct links in the lattice and hence transform differently under gauge transformation. Thus these supersymmetries do not commute with gauge transformations. Indeed, the condition that the parameters $(\delta\kappa_0, \delta\kappa_a, \delta\kappa_{ab})$ should be singlets under gauge transformations singles out $\cQ$ as the only supercharge that be exactly preserved on the lattice. 

Below, we show the supersymmetry transformations associated with $\cQ_a$ and $\cQ_{ab}$ supercharges. To emphasize the differing link characters of the fields involved we have adopted the expanded notation where a lattice field $f_a$ associated with link $\bn\to \bn+\hatbmu_a$ is denoted $f_a(\bn,\bn+\hatbmu_a)$
\bea
\label{eq:qa-ub}
\cQ_a \cU_b(\bn, \bn+\hatbmu_b) &=& \hf \delta_{ab}\eta(\bn, \bn), \\
\cQ_a \cUb_b(\bn+\hatbmu_b, \bn) &=& -\chi_{ab}(\bn+\hatbmu_a+\hatbmu_b, \bn), \\
\cQ_a \psi_b(\bn, \bn+\hatbmu_b) &=& \hf \delta_{ab} d_a(\bn, \bn) + (1-\delta_{ab})(\cDb^{(+)}_a\cU_b)(\bn, \bn+\hatbmu_a+\hatbmu_b), \\
\cQ_a \chi_{bc}(\bn+\hatbmu_b+\hatbmu_c, \bn) &=& -\hf \epsilon_{abcgh} (\cD_g^{(+)}\cU_h)(\bn, \bn+\hatbmu_g+\hatbmu_h), \\
\cQ_a \eta(\bn, \bn) &=& 0, \\
\label{eq:qa-d}
\cQ_a d_a(\bn, \bn) &=& 0.
\eea

\bea
\cQ_{ab} \cU_c(\bn, \bn+\hatbmu_c) &=& \hf \epsilon_{abcgh}\chi_{gh}(\bn+\hatbmu_g+\hatbmu_h, \bn), \\
\cQ_{ab} \cUb_c(\bn+\hatbmu_c, \bn) &=& \delta_{ac}\psi_b(\bn, \bn+\hatbmu_b) - \delta_{bc}\psi_a(\bn, \bn+\hatbmu_a), \\
\cQ_{ab}\psi_c(\bn, \bn+\hatbmu_c) &=& \epsilon_{abcgh}(\cDb_g^{(+)} \cUb_h)(\bn+\hatbmu_g+\hatbmu_h, \bn), \\
\cQ_{ab}\chi_{cd}(\bn+\hatbmu_c+\hatbmu_d) &=& \hf \delta_{ac}\delta_{bd} d_{ab}(\bn, \bn) - \delta_{bc}(\cD_a^{(+)}\cUb_d)(\bn, \bn) - \delta_{ac}\cD_b^{(+)}\cUb_d(\bn, \bn), \\
\cQ_{ab} \eta(\bn, \bn) &=& 2(\cD_a^{(+)}\cU_b)(\bn, \bn+\hatbmu_a+\hatbmu_b), \\
\cQ_{ab} d_{ab}(\bn, \bn) &=& 0.
\eea
Of course the translation between continuum and lattice expressions is not unique; one can always change the arguments of the lattice fields by terms of order the lattice spacing and add terms which are subleading in the lattice spacing to the supersymmetry transformations. However, this cannot change the essential result that any supersymmetry transformation other than that associated with the scalar $\cQ$ will necessarily relate lattice objects of differing link character and hence these supersymmetries will {\it not} commute with gauge transformations. This has an immediate consequence; the variation of any gauge invariant closed loop under either $\cQ_a$ or $\cQ_{ab}$ will lead to a gauge variant quantity whose expectation value will vanish by Elitzur's theorem \cite{Elitzur:1975im}. Thus Ward identities constructed this way will be satisfied trivially because of gauge invariance. Indeed, the only way to construct non-vanishing supersymmetry variations is to start from a non-gauge invariant operator corresponding to an open loop which closes under supersymmetry variation. We have been able to construct only a extremely small number of such operators and they are all of short length. This hints perhaps that the supersymmetric Ward identities associated with the non-scalar twisted supersymmetries supply very few additional constraints on the structure of the quantum theory over and above gauge invariance.

%###################################################################
\section{Discrete R-symmetries and the lattice theory}
\label{sec:Rsym-and-latt}
%###################################################################

In spite of the fact that the lattice action has the same form as 
the continuum theory, with a straightforward transcription of fields 
and covariant derivatives, we will show below that it does not enjoy the discrete R-symmetries 
$R_a$ and $R_{ab}$ that were enumerated above. Thus we cannot draw the 
same implication as in the continuum, that under renormalization (e.g., 
in the determination of the low energy effective action under Wilsonian 
renormalization group flow), the coefficients $\alpha_i$ that were discussed
above will all be equal. However, if any of the 15 discrete R-symmetries 
emerge in the long distance theory, this equality is sufficient to yield the 
full $\cN = 4$ supersymmetry at low energies.  Thus apart from a 
renormalization of the overall coefficient, which is just the inverse 
gauge coupling squared, no fine-tuning of the lattice action would in this case be 
required. 

There is some reason to hope that the discrete R-symmetries 
do in fact emerge at low energies.  The reason is that the twisting process 
has combined the $SO(4)$ spacetime symmetry with an $SO(4)$ subgroup of the 
R-symmetry group. If the $SO(4)$ rotational symmetry of the 
continuum emerges at low energies, then we also expect to obtain 
the $SO(4)$ subgroup of R-symmetry, since they are basically
on the same footing as far as twisting is concerned.  In that case some of the 
discrete R-symmetries would also be emergent, which is sufficient 
to guarantee the equality of coefficients.

To see why the discrete R-symmetries are not invariances of the lattice theory, note that they relate site fields to link fields and vice versa. This breaks the gauge invariance of the lattice theory: closed loops will transform to expressions that are no longer closed loops. As an example consider the following choice for $R_a$:
\bea
\label{eq:eta-psia-latt}
\eta(\bn, \bn) &\rightarrow& 2\psi_a(\bn, \bn + \hatbmu_a), \\
\psi_a(\bn, \bn + \hatbmu_a) &\rightarrow& \hf \eta(\bn, \bn), \\
\psi_b(\bn, \bn + \hatbmu_b) &\rightarrow& -\chi_{ab}(\bn + \hatbmu_a + \hatbmu_b, \bn), \\
\chi_{ab}(\bn + \hatbmu_a + \hatbmu_b, \bn) &\rightarrow& -\psi_b(\bn, \bn + \hatbmu_b), \\
\chi_{bc}(\bn + \hatbmu_b + \hatbmu_c, \bn) &\rightarrow& \hf \epsilon_{bcagh}\chi_{gh}(\bn + \hatbmu_g + \hatbmu_h, \bn),
\eea
and
\bea
\cU_a(\bn, \bn + \hatbmu_a) &\rightarrow& \cU_a(\bn, \bn + \hatbmu_a),~~\cU^\dagger_a(\bn + \hatbmu_a, \bn) \rightarrow \cU^\dagger_a(\bn + \hatbmu_a, \bn), \\
\label{eq:db-dbbar-latt}
\cU_b(\bn, \bn + \hatbmu_b) &\rightarrow& \cU^\dagger_b(\bn + \hatbmu_b, \bn),~~\cU^\dagger_b(\bn + \hatbmu_b, \bn) \rightarrow \cU_b(\bn, \bn + \hatbmu_b),~~b \neq a.
\eea

Then one of the terms in our lattice action is
\bea
S_4 &=& - \alpha_2 \sum_{\bn, m} \Tr \eta(\bn, \bn) \Big(\psi_m(\bn, \bn + \hatbmu_m) \cU^{\dagger}_m(\bn + \hatbmu_m, \bn) \nn \\
&&- \cU^{\dagger}_m(\bn, \bn - \hatbmu_m)\psi_m(\bn - \hatbmu_m, \bn)\Big).
\eea

After exposing the index $a$
\bea
S_4 &=& - \alpha_2 \sum_{\bn} \Tr \eta(\bn, \bn) \Big(\psi_a(\bn, \bn + \hatbmu_a) \cU^{\dagger}_a(\bn + \hatbmu_a, \bn) \nn \\
&&- \cU^{\dagger}_a(\bn, \bn - \hatbmu_a)\psi_a(\bn - \hatbmu_a, \bn)\Big) \nn \\
&&- \alpha_2 \sum_{\bn, b} \Tr \eta(\bn, \bn) \Big(\psi_b(\bn, \bn + \hatbmu_b) \cU^{\dagger}_b(\bn + \hatbmu_b, \bn) \nn \\
&&- \cU^{\dagger}_b(\bn, \bn - \hatbmu_b) \psi_b(\bn - \hatbmu_b, \bn)\Big).
\eea

Applying the field interchanges
\bea
S_4 &\rightarrow& S_4' = - \alpha_2 \sum_{\bn} \Tr 2 \psi_a(\bn, \bn + \hatbmu_a) \Big(\hf \eta(\bn, \bn) \cU^{\dagger}_a(\bn + \hatbmu_a, \bn) \nn \\
&&- \cU^{\dagger}_a(\bn, \bn - \hatbmu_a) \hf \eta(\bn - \hatbmu_a, \bn - \hatbmu_a) \Big) \nn \\
&&- \alpha_2 \sum_{\bn, b} \Tr 2 \psi_a(\bn, \bn + \hatbmu_a) \Big(-\chi_{ab}(\bn + \hatbmu_a + \hatbmu_b, \bn) \cU_b(\bn, \bn + \hatbmu_b) \nn \\
&&+ \cU_b(\bn - \hatbmu_b, \bn) \chi_{ab}(\bn + \hatbmu_a, \bn - \hatbmu_b) \Big),
\eea
we see that none of the terms closed loops on the lattice, thus breaking gauge invariance. One can make alternate choices for $R_a$ that will render one of the terms a closed loop, but then none of the other terms will be closed loops. By these sorts of arguments it is easy to convince oneself that no definition of $R_a$ or $R_{ab}$ will be symmetries of the lattice theory.

%########################
\section{Conclusions}
\label{sec:conclusions}
%########################

In this article we have derived the full set of supersymmetry transformations 
relevant to the Marcus twist of $\cN = 4$ super Yang-Mills. The recent supersymmetric 
lattice construction of $\cN = 4$ super Yang-Mills is based on a discretization of 
this twisted theory in which just the scalar supersymmetry is preserved in the 
lattice theory while the other fifteen supersymmetries are broken by lattice effects
at a generic point in parameter space.
We have shown how to derive the explicit form of these additional 
supersymmetries by combining the scalar supersymmetry with a series of 
discrete R-symmetries. 

In a previous paper \cite{Catterall:2011pd} we have given general arguments, 
based on exact lattice symmetries, that the restoration of these additional 
supersymmetries may potentially require the tuning of three independent marginal 
couplings in the classical lattice action.  However, in this paper we find yet 
another remarkable property of the $\cQ$-exact twisted formulation:  if any one of
the 15 discrete R-symmetries are emergent in the infrared theory, 
the full supersymmetry of the target $\cN = 4$ theory is recovered 
{\it without any additional fine tuning}.  Since Lorentz invariance implies
at least $\cN=1$ supersymmetry, it is quite reasonable that a $U(1)_R$
symmetry or a discrete subgroup would emerge in the 
Lorentz invariant continuum limit.  In that case we
would expect one or more of the 15 discrete R-symmetries to be present.
Furthermore, we have argued that if the $SO(4)$ Euclidean (twisted) 
rotational invariance is emergent in the low energy theory, as is necessary 
to obtain a Lorentz invariant continuum limit, then the discrete R-symmetries 
will most likely be in force.  This is because the twisting process
treats the $SO(4)$ spacetime symmetry and the $SO(4)$ R-symmetry
in an even-handed way, so we expect that restoration of one
will imply restoration of the other.  However, this is not a
rigorous statement.  However, if it is true, the crucial question of supersymmetry restoration 
is determined by whether or not Lorentz invariance is automatic in the continuum limit.
Even if it is not true, checking for the restoration of the
discrete R-symmetries is quite a bit easier than attempting to
check SUSY WIs, because of the issue of operator mixing in the latter case.

Clearly the question of the restoration of the $SO(4)$ spacetime symmetry 
has straightforward tests in term of measurements 
of correlation functions. The $S_5$ symmetry group of the $A_4^*$ lattice 
goes a long way toward ensuring the desired limit.   For instance,
the bosonic term that appear in Eq.~(\ref{ldefa}) will give a Lorentz
invariant continuum limit in that sector.  Unfortunately, the
fermionic terms that appear in Eq.~(\ref{ldefa}) require $\alpha_1 = \alpha_2
= \alpha_4$ in order for Lorentz invariance to emerge.  Thus, a more 
detailed study of the quantum continuum limit in this theory needs 
to be performed both perturbatively and nonperturbatively, in order 
to understand the fate of Lorentz invariance and the discrete R-symmetries.
The fermions that we are using are equivalent to staggered fermions, and so restoration
of Lorentz invariance would proceed as it does in lattice QCD if we had the
full hypercubic symmetry group.  However, $S_5$ is less restrictive,
as is seen from the independent coefficients on the fermionic terms in Eq.~(\ref{ldefa}).

Thus we see that it is not really a matter of there being {\it new} operators that break the discrete R-symmetry or Lorentz symmetry, since the operators in Eq. (\ref{ldefa}) are the same as in the bare action.  As we have just stated, in order to have Lorentz symmetry, the coefficients of all the fermion terms have to be the same, $\alpha_1 = \alpha_2 = \alpha_4$ and to have in addition the discrete R-symmetries, one also requires $\alpha_3$ to match these coefficients. We are not aware of any mechanism that would enforce equality of these coefficients under renormalization group flow, so generically we expect logarithmic violations of both symmetries. However, whether or not this generic situation (which is the worst case scenario) actually occurs can only be determined by perturbative and/or nonperturbative calculations that need to be performed.  In the case of perturbative calculations we have already checked at one-loop in Ref. \cite{Catterall:2011pd} and found equality of the coefficients. To compute the same effects at two loops in lattice perturbation theory is notoriously difficult, because of the integrals which must be evaluated numerically. We think that it will be quite a bit easier to check the discrete R-symmetries by computing correlation functions with Monte Carlo simulations, which is what we are currently planning to do as the next step, based on the results of this paper.

Even if a positive answer to this question is obtained, which would indeed be remarkable, there would still be important items to be done. One thing is that irrelevant operators may break the supersymmetry by $\cO(a)$ effects, where $a$ is the lattice spacing. Thus it is still necessary to take the continuum limit. In a theory that is supposed to be conformal at all scales, taking the continuum limit is quite a bit different from in a QCD-like theory where one has asymptotic freedom and dimensional transmutation. That is, vanishing bare coupling does not correspond to the continuum limit in the lattice theory that we are studying. Rather, what one must do is study physics at scales $x$ much larger than the lattice spacing, which amounts to long distance correlation functions. It still needs to be studied how large $x$ must be relative to $a$ in order to obtain the continuum physics. How would we know if we are obtaining the continuum physics, considering that the results are nonperturbative? It is here that it is crucial that we know that the restoration of $SO(4)$ Lorentz invariance and $SU(4)_R$ symmetry, together with the exact $\cQ$ supersymmetry, is sufficient.  These are properties that can be checked through the measurement of correlation functions, which should be related to each other through the symmetries.

%############################
\acknowledgments
%############################

This work was supported in part by the U.S. Department of Energy grant under Contract No. DE-FG02-85ER40237 (SC) and DE-FG02-08ER41575 (JG) and the LDRD program at the Los Alamos National Laboratory (AJ).  SC and JG also acknowledge the hospitality of the Aspen Center for Physics, which is supported by the National Science Foundation Grant No. PHY-1066293, where some of this work was conducted.


\begin{thebibliography}{1}

\bibitem{Kaplan:2003uh} 
  D.~B.~Kaplan,
  ``Recent developments in lattice supersymmetry,''
  Nucl.\ Phys.\ Proc.\ Suppl.\  {\bf 129}, 109 (2004)
  [hep-lat/0309099].

\bibitem{Giedt:2006pd}
  J.~Giedt,
  ``Deconstruction and other approaches to supersymmetric lattice field theories,''
  Int.\ J.\ Mod.\ Phys.\  A {\bf 21}, 3039 (2006)
  [arXiv:hep-lat/0602007].

\bibitem{Catterall:2009it}
  S.~Catterall, D.~B.~Kaplan, M.~Unsal,
  ``Exact lattice supersymmetry,''
  Phys.\ Rept.\  {\bf 484}, 71-130 (2009).
  [arXiv:0903.4881 [hep-lat]].

\bibitem{Joseph:2011xy} 
  A.~Joseph,
  ``Supersymmetric Yang-Mills theories with exact supersymmetry on the lattice,''
  Int.\ J.\ Mod.\ Phys.\ A {\bf 26}, 5057 (2011)
  [arXiv:1110.5983 [hep-lat]].

\bibitem{Unsal:2006qp}
  M.~Unsal,
  ``Twisted supersymmetric gauge theories and orbifold lattices,''
  JHEP {\bf 0610}, 089 (2006).
  [hep-th/0603046].

\bibitem{Catterall:2007kn}
  S.~Catterall,
  ``From Twisted Supersymmetry to Orbifold Lattices,''
  JHEP {\bf 0801}, 048 (2008).
  [arXiv:0712.2532 [hep-th]].

\bibitem{Kaplan:2002wv}
  D.~B.~Kaplan, E.~Katz, M.~Unsal,
  ``Supersymmetry on a spatial lattice,''
  JHEP {\bf 0305}, 037 (2003).
  [hep-lat/0206019].
  
\bibitem{Cohen:2003xe}
  A.~G.~Cohen, D.~B.~Kaplan, E.~Katz, M.~Unsal,
  ``Supersymmetry on a Euclidean space-time lattice. 1. A Target theory with four supercharges,''
  JHEP {\bf 0308}, 024 (2003).
  [hep-lat/0302017].

\bibitem{Cohen:2003qw}
  A.~G.~Cohen, D.~B.~Kaplan, E.~Katz, M.~Unsal,
  ``Supersymmetry on a Euclidean space-time lattice. 2. Target theories with eight supercharges,''
  JHEP {\bf 0312}, 031 (2003).
  [hep-lat/0307012].

\bibitem{Kaplan:2005ta}
  D.~B.~Kaplan, M.~Unsal,
  ``A Euclidean lattice construction of supersymmetric Yang-Mills theories with sixteen supercharges,''
  JHEP {\bf 0509}, 042 (2005).
  [hep-lat/0503039].

\bibitem{Catterall:2003wd}
  S.~Catterall,
  ``Lattice supersymmetry and topological field theory,''
  JHEP {\bf 0305}, 038 (2003).
  [hep-lat/0301028].

\bibitem{Catterall:2001fr}
  S.~Catterall, S.~Karamov,
  ``Exact lattice supersymmetry: The Two-dimensional N=2 Wess-Zumino model,''
  Phys.\ Rev.\  {\bf D65}, 094501 (2002).
  [hep-lat/0108024].

\bibitem{Elliott:2008jp}
  J.~W.~Elliott, J.~Giedt and G.~D.~Moore,
  ``Lattice four-dimensional N=4 SYM is practical,''
  Phys.\ Rev.\  D {\bf 78} (2008) 081701
  [arXiv:0806.0013 [hep-lat]].

\bibitem{Catterall:2012yq} 
  S.~Catterall, P.~H.~Damgaard, T.~Degrand, R.~Galvez and D.~Mehta,
  ``Phase Structure of Lattice N=4 Super Yang-Mills,''
  JHEP {\bf 1211}, 072 (2012)
  [arXiv:1209.5285 [hep-lat]].

\bibitem{Catterall:2011aa} 
  S.~Catterall, R.~Galvez, A.~Joseph and D.~Mehta,
  ``On the sign problem in 2D lattice super Yang-Mills,''
  JHEP {\bf 1201}, 108 (2012)
  [arXiv:1112.3588 [hep-lat]].
  
\bibitem{Mehta:2011ud} 
  D.~Mehta, S.~Catterall, R.~Galvez and A.~Joseph,
  ``Supersymmetric gauge theories on the lattice: Pfaffian phases and the Neuberger 0/0 problem,''
  PoS LATTICE {\bf 2011}, 078 (2011)
  [arXiv:1112.5413 [hep-lat]].
 
\bibitem{Galvez:2012sv} 
  R.~Galvez, S.~Catterall, A.~Joseph and D.~Mehta,
  ``Investigating the sign problem for two-dimensional $\mathcal{N}=(2,2)$ and $\mathcal{N}=(8,8)$ lattice super Yang--Mills theories,''
  PoS LATTICE {\bf 2011}, 064 (2011)
  [arXiv:1201.1924 [hep-lat]].  

\bibitem{Ishii:2008ib} 
  T.~Ishii, G.~Ishiki, S.~Shimasaki and A.~Tsuchiya,
  ``N=4 Super Yang-Mills from the Plane Wave Matrix Model,''
  Phys.\ Rev.\ D {\bf 78}, 106001 (2008)
  [arXiv:0807.2352 [hep-th]].

\bibitem{Ishiki:2008te}
  G.~Ishiki, S.~W.~Kim, J.~Nishimura and A.~Tsuchiya,
  ``Deconfinement phase transition in N=4 super Yang-Mills theory on $R \times S^3$ from supersymmetric matrix quantum mechanics,''
  Phys.\ Rev.\ Lett.\  {\bf 102} (2009) 111601
  [arXiv:0810.2884 [hep-th]].

\bibitem{Nishimura:2009xm} 
  J.~Nishimura,
  ``Non-lattice simulation of supersymmetric gauge theories as a probe to quantum black holes and strings,''
  PoS LAT {\bf 2009}, 016 (2009)
  [arXiv:0912.0327 [hep-lat]].

\bibitem{Ishiki:2009sg}
  G.~Ishiki, S.~W.~Kim, J.~Nishimura and A.~Tsuchiya,
  ``Testing a novel large-$N$ reduction for $\cN=4$ super Yang-Mills theory on $R \times S^3$,''
  JHEP {\bf 0909} (2009) 029
  [arXiv:0907.1488 [hep-th]].

\bibitem{Honda:2010nx} 
  M.~Honda, G.~Ishiki, S.~-W.~Kim, J.~Nishimura and A.~Tsuchiya,
  ``Supersymmetry non-renormalization theorem from a computer and the AdS/CFT correspondence,''
  PoS LATTICE {\bf 2010}, 253 (2010)
  [arXiv:1011.3904 [hep-lat]].

\bibitem{Hanada:2010kt} 
  M.~Hanada, S.~Matsuura and F.~Sugino,
  ``Two-dimensional lattice for four-dimensional N=4 supersymmetric Yang-Mills,''
  Prog.\ Theor.\ Phys.\  {\bf 126}, 597 (2011)
  [arXiv:1004.5513 [hep-lat]].

\bibitem{Hanada:2010gs} 
  M.~Hanada,
  ``A proposal of a fine tuning free formulation of 4d N = 4 super Yang-Mills,''
  JHEP {\bf 1011}, 112 (2010)
  [arXiv:1009.0901 [hep-lat]].

\bibitem{Sugino:2003yb}
  F.~Sugino,
  ``A Lattice formulation of superYang-Mills theories with exact supersymmetry,''
  JHEP {\bf 0401} (2004) 015
  [hep-lat/0311021].

\bibitem{Sugino:2004qd} 
  F.~Sugino,
  ``SuperYang-Mills theories on the two-dimensional lattice with exact supersymmetry,''
  JHEP {\bf 0403}, 067 (2004)
  [hep-lat/0401017].

\bibitem{Kanamori:2008bk} 
  I.~Kanamori and H.~Suzuki,
  ``Restoration of supersymmetry on the lattice: Two-dimensional N = (2,2) supersymmetric Yang-Mills theory,''
  Nucl.\ Phys.\ B {\bf 811}, 420 (2009)
  [arXiv:0809.2856 [hep-lat]].

\bibitem{Hanada:2009hq} 
  M.~Hanada and I.~Kanamori,
  ``Lattice study of two-dimensional N=(2,2) super Yang-Mills at large-N,''
  Phys.\ Rev.\ D {\bf 80}, 065014 (2009)
  [arXiv:0907.4966 [hep-lat]].

\bibitem{Hanada:2010qg} 
  M.~Hanada and I.~Kanamori,
  ``Absence of sign problem in two-dimensional N = (2,2) super Yang-Mills on lattice,''
  JHEP {\bf 1101}, 058 (2011)
  [arXiv:1010.2948 [hep-lat]].


\bibitem{Farchioni:2001wx} 
  F.~Farchioni {\it et al.}  [DESY-Munster-Roma Collaboration],
  ``The Supersymmetric Ward identities on the lattice,''
  Eur.\ Phys.\ J.\ C {\bf 23}, 719 (2002)
  [hep-lat/0111008].

\bibitem{Marcus:1995mq}
  N.~Marcus,
  ``The Other topological twisting of N=4 Yang-Mills,''
  Nucl.\ Phys.\  {\bf B452}, 331-345 (1995).
  [hep-th/9506002].

\bibitem{Kapustin:2006pk} 
  A.~Kapustin and E.~Witten,
  ``Electric-Magnetic Duality And The Geometric Langlands Program,''
  Commun.\ Num.\ Theor.\ Phys.\  {\bf 1}, 1 (2007)
  [hep-th/0604151].
  
\bibitem{D'Adda:2005zk} 
  A.~D'Adda, I.~Kanamori, N.~Kawamoto and K.~Nagata,
  ``Exact extended supersymmetry on a lattice: Twisted N=2 super Yang-Mills in two dimensions,''
  Phys.\ Lett.\ B {\bf 633}, 645 (2006)
  [hep-lat/0507029]. 

\bibitem{Catterall:2011pd} 
  S.~Catterall, E.~Dzienkowski, J.~Giedt, A.~Joseph and R.~Wells,
  ``Perturbative renormalization of lattice N=4 super Yang-Mills theory,''
  JHEP {\bf 1104}, 074 (2011)
  [arXiv:1102.1725 [hep-th]].

\bibitem{Elitzur:1975im} 
  S.~Elitzur,
  ``Impossibility of Spontaneously Breaking Local Symmetries,''
  Phys.\ Rev.\ D {\bf 12}, 3978 (1975).

\end{thebibliography}
\end{document}